\documentclass[letterpaper]{article} 
\usepackage{aaai2026}  
\usepackage{times}  
\usepackage{helvet}  
\usepackage{courier}  
\usepackage[hyphens]{url}  
\usepackage{graphicx} 
\usepackage{natbib}  
\usepackage{caption} 
\usepackage{algorithm}
\usepackage{algorithmic}
\usepackage{amssymb}
\usepackage{amsmath}
\usepackage{booktabs}
\usepackage{newfloat}
\usepackage{listings}
\DeclareCaptionStyle{ruled}{labelfont=normalfont,labelsep=colon,strut=off} 
\floatstyle{ruled}
\newfloat{listing}{tb}{lst}{}
\floatname{listing}{Listing}
\title{PhenoMoler: Phenotype-Guided Molecular Optimization via Chemistry Large Language Model}
\author{
    Ran Song, Hui Liu*
}
\affiliations{
    College of Computer and Information Engineering, Nanjing Tech University, Nanjing, 211800, China.\\

    hliu@njtech.edu.cn
}

\usepackage{bibentry}

\begin{document}

\maketitle

\begin{abstract}
Current molecular generative models primarily focus on improving drug–target binding affinity and specificity, often neglecting the system-level phenotypic effects elicited by compounds. Transcriptional profiles, as molecule-level readouts of drug-induced phenotypic shifts, offer a powerful opportunity to guide molecular design in a phenotype-aware manner. We present PhenoMoler, a phenotype-guided molecular generation framework that integrates a chemistry large language model with expression profiles to enable biologically informed drug design. By conditioning the generation on drug-induced differential expression signatures, PhenoMoler explicitly links transcriptional responses to chemical structure. By selectively masking and reconstructing specific substructures—scaffolds, side chains, or linkers—PhenoMoler supports fine-grained, controllable molecular optimization. Extensive experiments demonstrate that PhenoMoler generates chemically valid, novel, and diverse molecules aligned with desired phenotypic profiles. Compared to FDA-approved drugs, the generated compounds exhibit comparable or enhanced drug-likeness (QED), optimized physicochemical properties, and superior binding affinity to key cancer targets. These findings highlight PhenoMoler's potential for phenotype-guided and structure-controllable molecular optimization. 
\end{abstract}


\section{Introduction}
Molecular design sits at the heart of drug development, aiming to generate compounds with precise bioactivity and ideal pharmacokinetic properties~\cite{ramallo2011chemically,perucca2003ideal}. As disease complexity increases and precision‑medicine demands grow, traditional high‑throughput screening struggles with cost and efficiency~\cite{mayr2008future}. Targeted refinement of known drugs can dramatically boost potency, selectivity, and bioavailability while reducing off‑target toxicity, offering a faster, more economical route to next‑generation therapeutics~\cite{lin2019off}.

In recent years, computational molecular generation has witnessed remarkable progress~\cite{meuwly2021machine}, promoted especially by deep learning-based  models~\cite{kallioras2020dzaiℕ}. These methods capture the nuanced relationships between chemical structures and molecular properties, enabling the generation of compounds with tailored traits. However, existing methods optimize solely for high affinity and target specificity, neglecting system-level phenotypic effects and often yielding molecules that scored well in silico but fail to elicit the desired bioactivity in complex cellular environments. Also, many methods lack fine‑grained control over specific molecular substructures—such as scaffolds, side chains, or linkers—hindering their applicability to complex drug‑design challenges that demand precise, customized optimization. On the other hand, transcriptional profiles offer insights into how compounds alter cellular states, reflecting the molecule-level readout of perturbative effect on higher‑order phenotypes~\cite{lamb2006connectivity}. Owing to their scalability and cost‑effectiveness, transcriptional responses are extensively employed as intermediate measure of causal pathways from genotype to disease phenotype. For example, differential expression analysis has proven in identifying compounds capable of driving desired phenotypic shifts~\cite{Subramanian2017L1000}. Consequently, we use the terms ``expression profile" and ``phenotypic profile" interchangeably throughout this study.

In this paper, we introduce PhenoMoler, a GPT-like generative framework that integrates chemistry large language model and phenotypic profiles for molecular optimization. We first pretrain anutoregressive language model to generate chemically valid molecules from scratch. A two-layer one-dimensional convolutional neural network (1D-CNN) are used to tokenize phenotypic profiles. During fine‑tuning, we selectively mask predefined chemical substructures—scaffolds, side chains, or linkers, and leverage the cross‑attention mechanism to fuse the tokenized phenotypic profiles and molecular contexts to guide the reconstruction of masked fragment. Once trained, PhenoMoler enables substructure-level optimization of chemically valid molecules. Our empirical experiments demonstrate that phenotypic profiles reliably guide the generation toward biologically relevant and syntactically valid molecules. To our best knowledge, this is the first work to perform substructure-level optimization explicitly guided by phenotypic profiles.

The main contributions of this study are summarized as below:
\begin{itemize}
    \item We present a multi-modal generative framework that integrates a chemistry large language model with conditioning on drug‑induced transcriptional response, enabling molecule design aligned with cellular phenotype.
    \item We introduce a substructure-wise mask‑and‑reconstruct mechanism for scaffolds, linkers, and side chains, providing fine‑grained, controllable generation of chemical motifs to boost diversity and interpretability.
    \item Extensive experiments has been performed to assess the model's ability to optimize molecular substructures under phenotype-conditioned settings.
\end{itemize}

\section{Related Works}
\subsection{Computational Molecular Optimization}
Traditional drug molecule design has long relied on the expertise of medicinal chemists and iterative "Design–Make–Test–Analyze" (DMTA) cycles~\cite{wesolowski2016strategies}. With the rapid advancement of computational chemistry and artificial intelligence (AI), computational molecular generation has emerged as a critical strategy for accelerating drug discovery. Existing methods can broadly be categorized into: generative model-based method~\cite{zhavoronkov2019deep}, reinforcement learning-based methods~\cite{wang2024reinforcement}, and hybrid or ensemble methods~\cite{ardabili2019advances}. Among these, generative models have become the dominant paradigm. For example, REINVENT~\cite{blaschke2020reinvent} and DrugEx~\cite{sicho2023drugex} utilize sequence-based generative frameworks, such as Recurrent Neural Networks (RNNs)~\cite{medsker2001recurrent} and Transformers~\cite{vaswani2017attention}, to generate novel molecular structures in SMILES or SELFIES formats. In parallel, graph generative models—such as JT-VAE~\cite{jin2018junction} that encode molecular graphs into continuous latent spaces, seek to generate molecules by navigating the latent manifold. Additionally, flow-based models (e.g., GraphAF~\cite{shi2020graphaf}) and diffusion models (e.g., G2D-Diff~\cite{kim2024accelerating}) have demonstrated promising performance in learning complex chemical distributions and generating high-quality molecules.

Reinforcement learning approaches optimize molecules by learning a sequence of actions to iteratively construct or modify molecular structures in order to maximize a predefined reward function. For instance, MolDQN~\cite{zhou2019optimization} formulates molecular construction as a Markov Decision Process (MDP) and applies Deep Q-Networks (DQNs) to learn effective policies.  Furthermore, GFlowNets~\cite{bengio2023gflownet} have been employed to efficiently sample from approximate Pareto frontiers in multi-objective optimization tasks. With the rise of large language models (LLMs), researchers have begun leveraging their generative and reasoning capabilities for molecular design. Models such as Chematica~\cite{grzybowski2018chematica} and Chemma~\cite{xu2022chemma} integrate strategies like genetic algorithms, rejection sampling, and prompt tuning, achieving state-of-the-art performance on various molecular optimization benchmarks. Despite these advancements, current approaches primarily focus on optimizing molecular properties such as binding affinity and target specificity, while overlooking the system-level phenotypic effects induced by compounds. Moreover, they often lack fine-grained control over specific molecular substructures, such as scaffolds, side chains, and linkers, which are critical for achieving precise and interpretable molecular optimization.

\subsection{Phenotypic Drug Generation}
Phenotype‑driven drug discovery traces back to the Connectivity Map~\cite{lamb2006connectivity}, matching drug‑induced expression signatures to disease states. The L1000 platform \cite{subramanian2017functional} scaled this paradigm, enabling transcriptome‑guided molecular generation. Conditional generative models have since emerged. For example, FAME~\cite{pham2022fame} first denoises the expression profiles and employs a fragment-based conditional variational autoencoder (VAE) that generates molecular fragments from phenotypic signatures to produce novel bioactive compounds. Gx2Mol~\cite{li2024gx2mol} and GxVAEs~\cite{li2024gxvaes} also utilize VAEs to embed expression profiles into latent spaces, which are then decoded to generate novel molecular structures. MolGene-E~\cite{ohlan2025molgene} harmonizes and denoises chemical-perturbed transcriptomes with a cross-modal VAE, then uses a CLIP-style contrastive prior to map gene signatures into chemical space and generate novel drug-like compounds. Recently, diffusion model has been used to generated molecules. G2D-Diff~\cite{kim2025genotype} uses a latent diffusion framework conditioned on somatic genotype and desired drug response to directly generate diverse, potency-aligned anticancer small molecules. MolEditRL~\cite{zhuang2025moleditrl} pretrains a discrete graph diffusion model to reconstruct molecules from source structures and text instructions, then fine-tunes with reinforcement learning to edit and optimize properties while preserving structural integrity. Despite progress, these methods often lack precise, substructure‑specific control or fail to jointly satisfy phenotypic efficacy and target binding. Our method bridges this gap by introducing substructure‑wise masking and a cross‑attention fusion of transcriptomic and chemical contexts to optimize reference molecule scaffolds, linkers, and side chains.

\begin{figure*}[ht]
        \centering
        \includegraphics[width=1\linewidth]{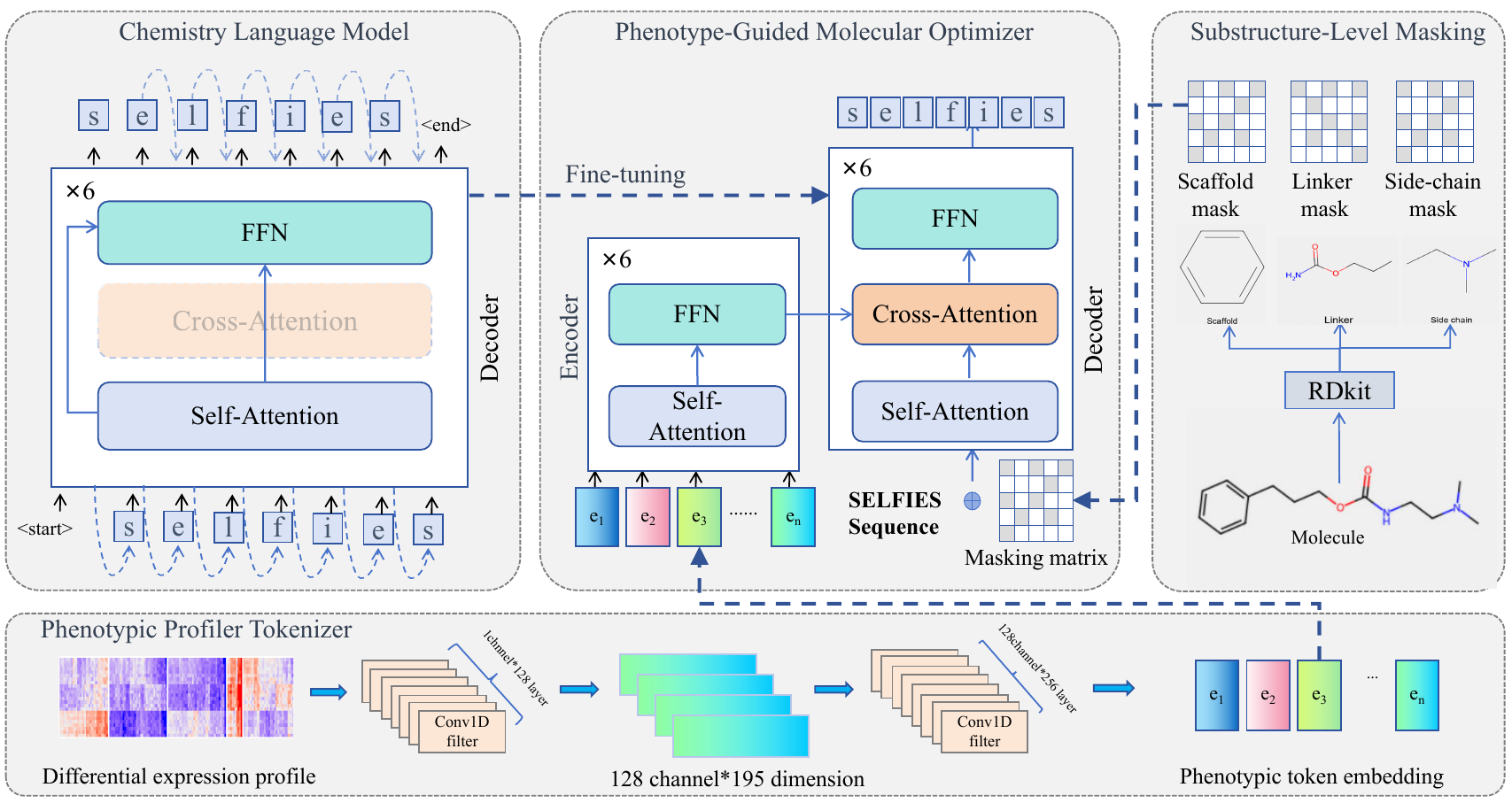}
        \caption{Illustration of PhenoMoler architecture. It comprises a pretrained large language model, a phenotype profile tokenizer based on 1D-CNN network, a chemical substructure masker, and a cross-attention-based molecular optimizer.}
        \label{fig:phenotype}
\end{figure*}

\section{Methods}
\subsection{Framework Overview}
We propose PhenoMoler, a phenotype-guided molecular generation framework built upon a chemistry large language model. As illustrated in Figure 1, the framework consists of three core components: a chemistry large language model, a phenotypic profile tokenizer, and phenotype-conditioned molecular optimizer. Formally, to explicitly modeling the molecule generation process conditioned on transcriptomic signatures, we establish a mapping function $f: \mathbb{R}^n \rightarrow \mathcal{Y}^T$, where $\mathbb{R}^n$ denotes the space of phenotypic profiles, and $\mathcal{Y}^T$ represents the space of chemical structures. We firstly pretrain an autoregressive language model using large-scale molecular sequences in SELFIES format. By learning the contextual dependencies and grammar of molecular sequences, it is able to generate chemically valid molecules from scratch. Next, we selectively mask chemical substructures (scaffolds, side chains, or linkers) of a molecule and then use drug-induced differential expression profiles as conditional input to reconstruct the molecules. Through a cross-attention mechanism, PhenoMoler integrates pharmacogenomic data and chemical context to reconstruct the masked fragments in a phenotype-guided manner. Our experimental results show that phenotypic signals effectively guide the recovery of structurally valid and biologically meaningful molecules.

\subsection{Chemistry Language Model}
To effectively learn molecular syntax and chemical rules, we pretrain an autoregressive language model based on Transformer decoder architecture using a large-scale set of bioactive compounds obtained from ZINC database. The decoder consists of six stacked standard Transformer layers, each composed of multi-head self-attention, feedforward networks, layer normalization, and residual connections. Compounds are represented using SELFIES strings, which enforce a formal grammar that guarantees chemically valid outputs and prevents syntactic or semantic errors. We fix the maximum sequence length to match the longest SELFIES molecule in the dataset, and pad shorter sequences with a special `[PAD]' token. Each token is embedded into a 256-dimensional vector and augmented with positional encoding to retain sequential information. To preserve the autoregressive nature of generation, we apply a causal mask in the self-attention layers to prevent the model from accessing future tokens during training.

Let $X\in \mathbb{R}^{T \times d}$ be an embedded input SELFIES sequence, where $T$ is the sequence length and $d$ is the embedding dimension. We define the linear projections of queries, keys, and values as $Q_X=XW^Q$, $K_X=XW^K$, and $V_X=XW^V$, where $W^Q, W^K, W^V \in \mathbb{R}^{d \times d_k}$ are trainable weight matrices, $d_k$ is the dimension of each attention head, and the self-attention mechanism at each layer operates as follows:
\begin{equation}
    \text{SelfAttn}(Q_X, K_X, V_X) = \text{softmax}\left( \frac{Q_XK_X^\top}{\sqrt{d_k}} + M \right)V_X
\end{equation}
in which $M$ is an upper-triangular mask matrix with $-\infty$ values used to block attention to future tokens, $\sqrt{d_k}$ serves as a scaling factor to prevent large dot-product values. 

\subsection{Phenotypic Profile Tokenizer}
The differential expression profiles should be converted to tokens as conditional inputs of molecular decoder. Since some genes often exhibit common expression patterns—where subsets of functionally related genes tend to be co-expressed or co-silenced in response to drug treatment—we propose a two-layer multi-channel one-dimensional convolutional neural network (1D-CNN) to capture these regulatory modules. Formally, the input is a one-dimensional vector that represents the differential expressions induced by drug perturbations. The first convolutional layer employs 128 filters of size 1$\times$5 with stride 5, performing sliding-window operations to extract local co-expression features. This results in a feature tensor of shape 128$\times$195, encoding 195 local receptive fields across 128 channels. The second convolutional layer further applies 256 filters of shape 128$\times$1$\times$3 with stride 2, aggregating the multi-channel features into higher-order representations. The final output is a compact phenotype embedding with dimensions 256$\times$97. As a result, the phenotypic tokenizer not only performs effective dimensionality transform, but also explicitly captures drug-induced gene co-expression patterns.



\subsection{Substructures-Level Masking}
To enable controllable molecular generation, we mask and reconstruct distinct molecular substructures guided by phenotypic profiles. Specifically, we decompose each molecule into three key components based on bond connectivity in its molecular graph: scaffold, linker, and side chains. The scaffold governs the overall conformation and pharmacophore placement, the linker modulates spatial arrangement and flexibility between functional motifs, and the side chains contribute to molecular diversity and tunability by influencing hydrophobicity, polarity, and binding affinity.

We use the RDKit tool to extract the Murcko scaffold~\cite{landrum2013rdkit}, then identify linkers as the shortest subgraphs connecting scaffold and substituents, and determine the side chains as non‑scaffold branches from those linkers. Next, we replace the SELFIES fragments corresponding to each substructure with special mask tokens, establishing three separate mask‑and‑reconstruct tasks (one per substructure type).  For each substructure type, we establish an independent mask-and-reconstruct task. Our model is trained to recover the masked fragments conditioned on (1) the remaining visible molecular context and (2) the differential expression profile induced by this molecule. By decoupling scaffold, linker, and side‑chain generation, our approach dramatically reduces the effective search space and avoids the inefficiencies of sampling entire molecules. This modular strategy enhances both generation efficiency and controllable generation under phenotypic guidance.


\begin{table*}[!htbp]
\centering
\begin{tabular}{|l|c|c|c|c|c|c|}
\toprule   \hline 
\textbf{Target} & \textbf{Generated} & \textbf{\#Exact Matches} & \textbf{Similarity (mean $\pm$ std)} & \textbf{Novelty (\%)} & \textbf{Uniqueness (\%)} & \textbf{Validity (\%)}   \\ \hline
\midrule
AKT1    & 41224  & 599  & 0.43 $\pm$ 0.02 & 96.0   & 100.0 & 95.86 \\ \hline
AKT2    & 23954  & 362  & 0.42 $\pm$ 0.02 & 85.1   & 100.0 & 95.34 \\ \hline
AURKB   & 31204  & 598  & 0.48 $\pm$ 0.03 & 81.6   & 100.0 & 96.74 \\ \hline
CTSK    & 19514  & 320  & 0.52 $\pm$ 0.05 & 84.16  & 100.0 & 94.48 \\ \hline
EGFR    & 81244  & 1152 & 0.453 $\pm$ 0.05 & 87.0  & 100.0 & 95.53 \\ \hline
HDAC1   & 33974  & 1054 & 0.59 $\pm$ 0.04 & 72.0   & 100.0 & 96.53 \\ \hline
MTOR    & 62654  & 1109 & 0.46 $\pm$ 0.02 & 84.21  & 100.0 & 91.18 \\ \hline
PIK3CA  & 45354  & 693  & 0.48 $\pm$ 0.03 & 85.14  & 100.0 & 86.53 \\ \hline
SMAD3   & 31601  & 409  & 0.39 $\pm$ 0.03 & 87.25  & 100.0 & 97.5  \\ \hline
TP53    & 321524 & 8375 & 0.53 $\pm$ 0.03 & 74.52  & 100.0 & 96.19 \\   \hline 
\bottomrule
\end{tabular}
\caption{ Evaluation of molecules generated via phenotype-guided scaffold optimization for ten well-characterized target proteins.}
\label{tab:drug_metrics}
\end{table*}

\subsection{Phenotype-Guided Molecular Optimizer}
We first employ a standard Transformer encoder to extract phenotype features. The tokenized phenotypic profiles by 1D-CNN network are fed into a 6-layer standard Transformer encoder, where each layer comprises multi-head self-attention, feed-forward networks, layer normalization, and residual connections. Next, to integrate phenotypic feature with molecular context, we introduce cross-attention layers into the pretrained decoder of the chemical language model. Specifically, the encoded phenotype embeddings are projected as keys and values, while the tokens of the molecular sequence act as queries. Formally, given the linear projection matrices $Q_X=XW^Q$, $K_G=GW^K$, $V_G=GW^V$, where $W^Q$are the learned weight matrices in pretraining stage, $X$ denotes the embedded molecule tokens, and $G$ denotes the extracted phenotype features, the cross-attention weights at each decoding layer are computed as:
\begin{equation}
    \text{CrossAttn}(Q_X, K_G,V_G) = \text{softmax}\left( \frac{Q_XK_G^T}{\sqrt{d}} \right) V_G
\end{equation}

This cross-attention structure enables the Transformer decoder to implement multi-level modality fusion, significantly enhancing the model's ability to align molecular generation with phenotypic effect. As a result, our model captures both the internal chemical dependencies and the external phenotypic regulatory cues, allowing for coherent reconstruction of the masked substructures  based on phenotype guidance. The objective is to recover the masked portion of the molecular sequence, we employ the cross-entropy loss as the optimization criterion, formally defined as:
\begin{equation}
\mathcal{L}_{\text{reco}}(\theta) = -\sum_{t \in \mathcal{T}} y_t \log P_{\theta}(y_t | y_{<t}, G)
\end{equation}
where $\mathcal{T}$ is the set of masked token, $y_t$ is the ground truth token at position $t$, and $G$ denotes the phenotype embedding. This loss encourages the model to generate  molecular fragments with chemically valid and phenotypic effect.


\begin{figure*}[t]
        \centering
        \includegraphics[width=0.95\linewidth]{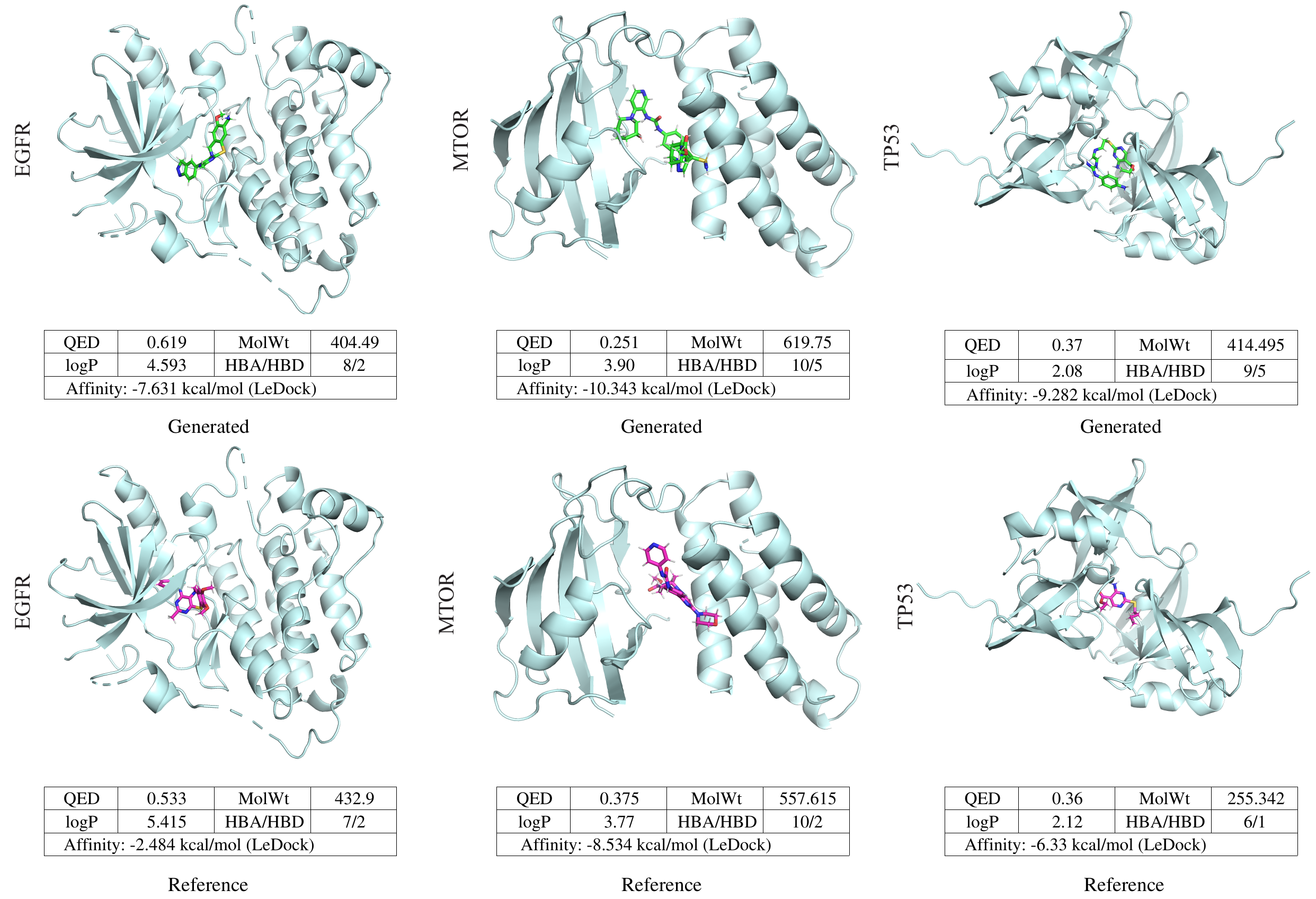}
        \caption{Comparison between reference ligands and PhenoMoler-generated molecules via scaffold optimization, evaluated in terms of binding affinity to target proteins and key physicochemical metrics. }
        \label{fig:phenotype}
\end{figure*}

\section{Experiments}
\subsection{Datasets}
We retrieve 10,032,879 bioactive molecular structures from ZINC database~\cite{irwin2005zinc} and then converted SMILES descriptors into SELFIES sequences. This large-scale molecule dataset was used to pretrain the decoder of chemistry large language model. The drug-induced expression profiles are obtained from L1000 project, which is a collective repository of transcriptional responses of human cell lines to drug exposures. We select the expression profiles of cell lines treated by 10$\mu$M drug concentration for 24 hours. To streamline the data, technical replicates were averaged. As a result, we obtained drug-induced expression profiles spanning 6,549 drugs and 164 cell lines. The refined dataset comprised a total of 86,400 expression profiles across 978 landmark genes. Since, we use Level5 (L5) data from the L1000 project, we derive differential expression signatures of each cell line under drug treatment by subtracting mean expression levels of control samples. The signatures serve as phenotypic profiles input to model, and the associated molecules in SELFIES format are masked by replacing substructures by [MASK] token and also taken as input for model fine-tuning. 

For performance evaluation, we focused on ten canonical oncogenic targets—AKT1, AKT2, AURKB, CTSK, EGFR, HDAC1, MTOR, PIK3CA, SMAD3, and TP53—each implicated in prevalent cancers. For each target, we curated a set of ligands experimentally validated ligands with confirmed binding affinity from the Drug Target Commons (DTC) database~\cite{tang2018drug} to serve as reference compounds. The differential expression signatures derived from genetic perturbations of corresponding genes in the MCF7 breast‑cancer cell line are used as phenotypic conditioning to generated new molecules. Finally, to assess the binding affinities of generated molecules, we retrieved high‑resolution three‑dimensional structures of all ten proteins from the Protein Data Bank (PDB)~\cite{burley2017protein}. For benchmarking, we systematically compare the performance metrics of the reference ligands and generated molecules to validate the effectiveness of our phenotype-guided generation strategy..

\begin{figure*}[ht]
        \centering
        \includegraphics[width=1\linewidth]{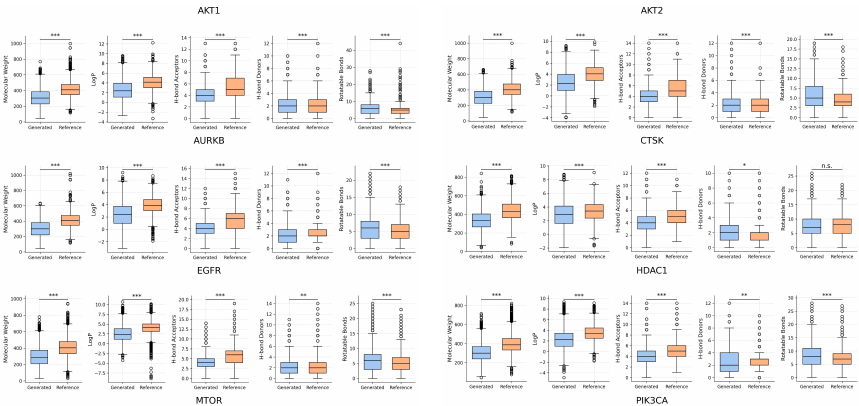}
        \caption{Comparison between PhenoMoler-generated molecules by side-chain optimization and reference ligands based on Lipinski's Rule of Five. Statistical significance is indicated by p$<$0.05 (*), p$<$0.01 (**), and p$<$0.001 (***).}
        \label{fig:phenotype}
\end{figure*}

\subsection{Performance Metrics}
For the generated molecules, we assessed their validity, uniqueness, novelty, and diversity. High validity and uniqueness are indicators of an effective molecule generation process, while high novelty and high diversity indicates the model is not overfitting to the training data. To assess the drug-likeness of the generated molecules, we employed the quantitative estimate of drug-likeness (QED) score. The topological polar surface area (TPSA) is also used to evaluate molecular polarity and passive membrane permeability. To evaluate the similarity between the generated molecules and known ligands of target proteins, we adopt Tanimoto similarity based on molecular fingerprints. Also, Lipinski’s Rule of Five—molecular weight (MolWt), octanol–water partition coefficient (LogP), numbers of H‑bond donors (HBD) and acceptors (HBA), and rotatable bond count—are used to gauge their physicochemical suitability for oral bioavailability. Moreover, we employed the docking scores computed by LeDock~\cite{ liu2019using} to assess the binding affinity to target proteins. 


\subsection{Phenotype-Guided Scaffold Optimization}
We fine‑tuned the pretrained chemistry language model on the scaffold mask-and-reconstruction task by using the differential expression signatures from L1000 dataset as phenotypic conditions. After training, we evaluated the model's performance as follows: for each of ten canonical targets, we selected bioactive drugs upon this target as references and supplied the MCF7‑derived differential expression signature alongside their masked scaffold as model input. We generated 50 candidate molecules for each reference compound, which we quantitatively assessed for optimization efficacy and examined for molecular diversity.

As shown in Table 1, for over 90\% of reference compounds, our model successfully generated at least one generated molecule whose Tanimoto similarity to the original scaffold is 1.0, indicating complete reconstruction of the masked scaffold. We further found that more than 80\% of the generated molecules exhibit novel scaffolds, and over 90\% are both syntactically and chemically valid (invalid molecules are excluded from Table 1), demonstrating the model's capability to produce valid and novel molecules. To further assess the quality of our generated compounds, we compared the QED distribution of model‑generated molecules against that of known ligands, visualizing the results with density plots for each protein target. As illustrated in Figure S1, the QED profiles of our generated compounds closely mirror those of known ligands. Notably, molecules generated for TP53 exhibit significantly higher QED values. These findings indicate that our phenotype‑guided generation process preserves pharmacological viability and yields structures of high drug‑likeness.

\begin{figure*}[ht]
        \centering        
        \includegraphics[width=0.95\linewidth]{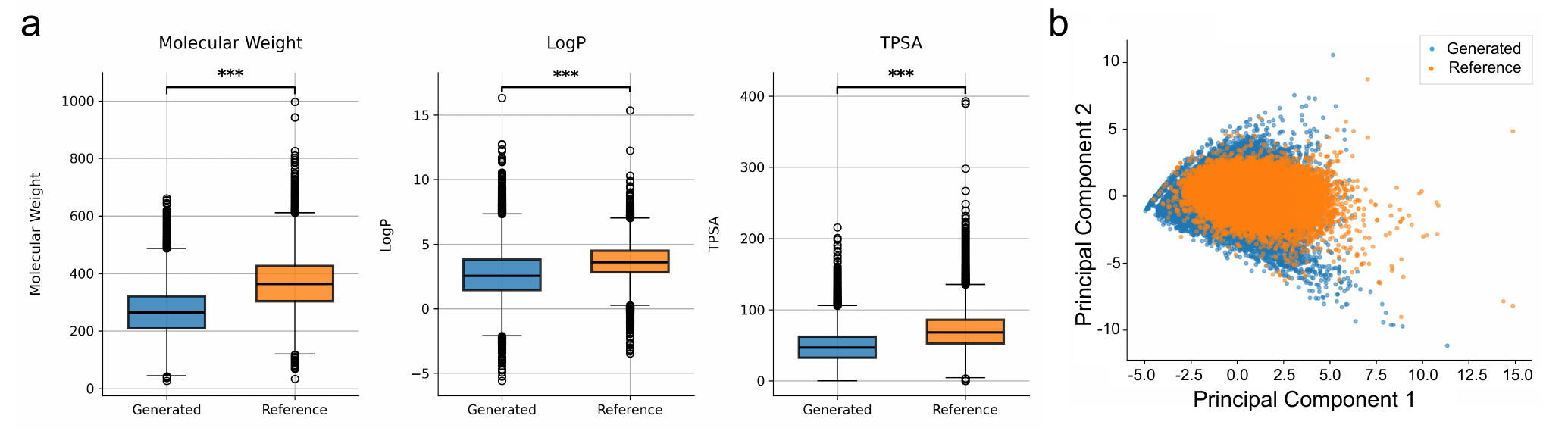}
        \caption{Comparison between PhenoMoler-generated molecules by linker optimization and reference ligands. Statistical significance is indicated by p$<$0.05 (*), p$<$0.01 (**), and p$<$0.001 (***).}
        \label{fig:phenotype}
\end{figure*}

We then examined molecules whose Tanimoto similarity to the reference drugs fell between 0.3 and 0.7, and observed that many of these candidates exhibited higher predicted binding affinities than the reference ones. Specifically, we used the LeDock docking tool~\cite{ LeDock docking tool} to dock each generated compound against its corresponding target protein. As shown in Figure 2, top-rank generated molecules acheved higher binding affinity to three typical disease targets (EGFR, mTOR and TP53) than reference ligands. Beyond binding affinity, the generated molecules also exhibit improved QED scores. These findings demonstrate that our phenotype‑guided scaffold generation process can effectively optimize molecular structure to yield novel compounds with superior target affinity.

\subsection{Side-Chain Optimization}
To further evaluate our model's performance, we devised a side‑chain optimization experiment that mimics the common ``side‑chain optimization" scenario in drug discovery—enhancing molecular properties while retaining the core scaffold. For benchmarking, we still selected reference compounds against the ten canonical targets and used their perturbation‑induced expression signatures in MCF7 breast‑cancer cell line as phenotypic guidance. The model was then tasked with reconstructing the masked side chains using the phenotypic signature and core scaffold as inputs, thereby enabling targeted side‑chain optimization.

For each reference ligands, we generated 50 candidates and evaluated them against Lipinski’s Rule of Five metrics (Figure 3, Figure S2). Compared to references, the candidates exhibited markedly lower molecular weights and cluster within the ideal 250–400Da range and thus improving pharmacokinetic property. Similarly, LogP values shifted into the optimal 1-3 range, reflecting a more balanced hydrophilicity–lipophilicity ratio. HBD and HBA counts remained within recommended ranges, with acceptors particularly trending toward the preferred 0–2 range to reduce off‑target binding and enhance selectivity and membrane permeability. Moreover, we conducted statistical significance tests between the reference and generated molecule groups. The results confirm that their differences are statistically significant (p-value$<$0.05, Wilcoxon rank sum test), indicating that the generated candidates achieve overall improvements across metrics. These results validate the efficacy and practical utility of the phenotype‑guided side‑chain optimization for enhancing drug‑like properties.

\subsection{Linker Optimization}
To evaluate the model's capacity in linker optimization, we conducted a linker‑masking and reconstruction experiment following the aforementioned workflow. Similarly, we generated 50 candidate molecules per reference ligand and evaluated three key physicochemical metrics: MolWt, LogP, and TPSA. As shown in Figure 4a, the mean MolWt of the candidates converge significantly on the $\sim$300Da ideal for small‑molecule drugs. We found that the median LogP values dropped, which indicates moderated lipophilicity and enhanced aqueous solubility supportive of improved ADME profiles. The TPSA values also declined compared to reference drugs, suggesting potential improvements in membrane permeability and intracellular uptake. Moreover, to confirm that the candidates remained within the drug‑like chemical space, we performed principal component analysis (PCA) on the molecular fingerprints and visualized them in principal component space (Figure 4b). The generated compounds and reference drugs exhibit strong overlap, while the candidates cover broader chemical region, demonstrating the model's ability to retain bioactivity features and explore novel chemically structural space.

\subsection{Therapeutic Molecular Generation}
To evaluate our model under patient phenotypic guidance, we selected three TCGA cases with transcriptomic profiles and their treatment records. TCGA‑QK‑A6IH (head and neck carcinoma) showed progression after paclitaxel monotherapy but responded partially to 5‑fluorouracil plus cetuximab. TCGA‑XF‑AAN7 (bladder carcinoma)  responded to gemcitabine but progressed following ifosfamide treatment. TCGA‑G2‑A3IE (bladder carcinoma) showed partial response to gemcitabine plus ifosfamide but later progressed under methotrexate, cisplatin, and vinblastine.

We obtained tumor RNA-seq profiles for three TCGA patients and matched normal tissue data from GTEx as controls. After Z-score normalization, differential expression signatures were computed by subtracting control means and then inverted to simulate therapeutic reversal by multiplying -1. These signatures served as phenotypic inputs for molecular optimization. For each patient, using their non-responsive drug as a reference, we performed scaffold and side-chain optimization, generating 30 candidate molecules per task (60 in total) for evaluation. We evaluated the Tanimoto similarity of each generated candidate to the effective drugs in comparison to  non‑responsive therapy. As illustrated in Figure S3, the majority of generated molecules achieve significantly higher similarity against effective drugs than against ineffective ones. The results indicate that our patient‑specific, phenotype‑guided generation strategy successfully shifts molecular generation toward chemical subspaces associated with proven clinical efficacy.

\section{Conclusion}
In this paper, we presented PhenoMoler, a phenotype-guided molecular generation framework that integrates gene expression signatures with chemistry language model. Experimental results demonstrate the generated candidates with enhanced drug-likeness, structural novelty, and target affinity compared to reference ligands.

\section*{Acknowledgments}
This work was supported by National Natural Science Foundation of China (No. 62372229), Natural Science Foundation of Jiangsu Province (No. BK20231271).

\bibliography{aaai2026}

\end{document}